\documentclass[twocolumn,aps,prl,showpacs,amssymb,raggedbottom,nobalancelastpage,superscriptaddress]{revtex4}

\usepackage{amsmath}
\usepackage{amssymb}
\usepackage{amsfonts}
\usepackage{dsfont}
\usepackage{graphicx}
\usepackage{bm}
\usepackage{color}
\usepackage{appendix}
\usepackage{epsfig}

\newcommand\be{\begin{equation}}
\newcommand\ee{\end{equation}}
\newcommand\bea{\begin{eqnarray}}
\newcommand\eea{\end{eqnarray}}

\begin{document}
\title{ Using weak measurements to extract the $Z_2$ index of a topological insulator}

\author{Zohar Ringel}
\affiliation{Theoretical Physics, Oxford University, 1, Keble Road, Oxford OX1 3NP, United Kingdom.}

\begin{abstract}
Recently there has been an interest in applying the concept of weak values and weak measurements to condensed matter systems. Here a weak measurement protocol is proposed for obtaining the $Z_2$ index of a topological insulator. The setup consists of a topological insulator with a hole pierced by a time dependent Aharonov-Bohm flux. A certain weak value ($A_{gs}$) associated with the time-integrated magnetization in the hole has a universal response to a small ambient magnetic field ($B$), namely $A_{gs}B = 2 \hbar$. This result is unaffected by disorder, interactions, and, to a large extent, the speed of the flux threading. It hinges mainly on preventing the flux from leaking outside the hole, as well as being able to detect magnetization at a resolution of a few spins. A similar result may be obtained using only charge measurements, in a setup consisting of a double quantum dot weakly coupled to an LC circuit. Here one obtains $\langle \phi \rangle_{weak} Q_0 =2\hbar$, where $\langle \phi \rangle_{weak}$ is a weak value associated with the flux on the inductor and $Q_0$ is the average capacitor charging. The universality of these results suggests that they may be used as a testbed for weak values in condensed matter physics. 
\end{abstract}


\maketitle

Topological insulators (TIs) have attracted much attention in recent years due to their novel bulk and surface properties~\cite{RMP_TI,RMP_TI2}. In their bulk, these materials are insulating and certain twists in the bulk's band structure are characterized by topological indices. This implies, via a bulk-edge correspondence, that the surfaces of these materials are metallic, and have a strong coupling between momentum and spin. Such profound spin dependent effects, that do not require external magnetic fields, are potentially useful in the field of spintronics \cite{Ilan2014,Wu2011,Ojeda2012}.

The topological character of a material may correspond to a quantized bulk response function, e.g.~, the quantized Hall conductance in the integer quantum Hall effect (QHE). This allows for the direct detection of the topological index of the material. However TIs, which in two dimensions (2D) can be pictured as two stacked QHE layers with opposite magnetic fields \cite{RMP_TI,RMP_TI2}, are not known to bear any quantized bulk response function or observable. As a result, experimental identification of a TI material is a more subtle task that relies on indirect evidence. For example analysis for surface ARPES spectrum \cite{RMP_TI,Hsieh2008} and measurement of edge conductance \cite{Konig2007}.  

Recently there has been an interest in applying a different measuring scheme, based on the idea of weak values \cite{Aharonov1988}, to condensed matter systems \cite{Romito2008, Oded2011,Romito2012,Blatter2014}. This measurement scheme allows one to measure off-diagonal matrix elements of operators directly, and hence extract more information than is available from a standard measurement \cite{Carmi2014}. Also, under certain circumstances, weak values can be used to amplify a weak signal \cite{Hosten2008,Oded2011}. A typical setup is a double quantum dot on which one applies various perturbation to induce either a Stueckelberg-Landau-Zener (Zener) transition \cite{Romito2012} or Rabi oscillations \cite{Romito2008} between two charge states of the device. The signal of a charge detector which is weakly coupled to the device, can then achieve values which exceed the classically allowed ones, provided that one post-selects only the measurements in which an unlikely outcome occurred. From an entirely different direction, a certain weak value (called the ``strange correlator'') has been used to identify power law correlations in symmetry protected topological phases \cite{Cenke2014}. However, measuring this weak value is unfeasible, as it would require waiting for an extremely unlikely event in which a quantum fluctuation in the topological insulator makes it appear as the ground state of a trivial insulator. In contrast, below we propose a more physical weak measurement which can be used to identify a TI.  

In a geometry with closed boundary conditions, a TI hosts a single unavoided Zener transition driven by Aharonov-Bohm (AB) fluxes \cite{Fu2006,Fu2007}. This transition occurs between the ground state and a magnetic excitation which resides on the boundary. Considering for instance a Corbino-disk geometry, the threading of a single AB flux quantum ($\phi_{0}$) through the hole in the disk results in a single level crossing, which occurs exactly when half the flux is threaded. The final state after the threading is orthogonal to the ground state and contains some magnetization \cite{magnetization}. A unique feature of this crossing point, is that only two levels which form a Kramers pair are involved \cite{Fu2006}. This ensures its persistence even as TRS respecting disorder and interactions are introduced \cite{Ringel2013,Ringel2014}. Moreover the orthogonality between the initial and final states remains unaltered also far away from the adiabatic limit, up to flux threading rates of roughly 1Teraherz \cite{speed, Ringel2013}. More generally a defining property of a TI (trivial band insulator) is that it hosts (does not host) such a transition \cite{Fu2006,Fu2007,many}. 


In this work, a measurement protocol of the $Z_2$ index of a topological insulator is proposed, which exploits the above Zener transition with its high degree of robustness. We consider a detector which is weakly coupled to the boundary magnetization and measure its signal during a threading of the AB flux. At the end of the threading, a regular (strong) magnetization measurement is performed to determine the final state. The weak detector signal is then conditionally averaged on having a non-magnetized final state. The result ($A_{gs}$), also known as a weak value (WV) \cite{Aharonov1988}, shows a quantized response to a small ambient magnetic field ($B_i$) which corresponds to the $Z_2$ invariant directly (see Eq. (\ref{eq-Ags})). Similarly quantized results are obtained for a double quantum dot coupled a quantum LC circuit, this time relating the weak value of the flux on the inductor $\langle \phi \rangle_{weak}$, with the charge on the capacitor $Q_0$. 

\begin{figure}[ht!]
\begin{center}
\includegraphics[width=\columnwidth]{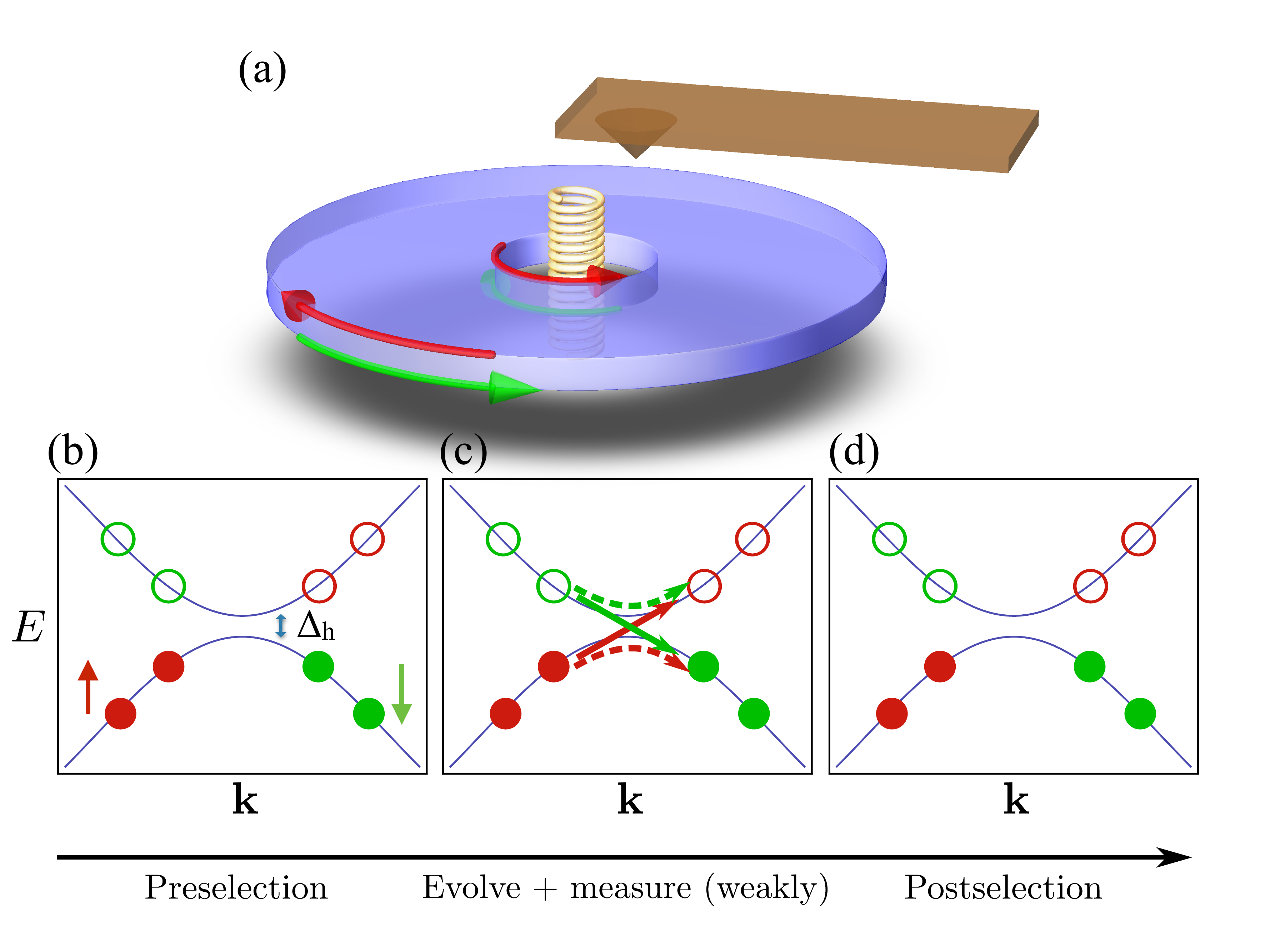}
\end{center}
\caption{(Color online) ({\bf a}) An illustration of a 2D topological insulator in a Corbino disk geometry. Each edge supports counter-propagating chiral modes of opposite spin (denoted by (green and red) counter-propagating arrows). Placing a solenoid within the disk allows for a time-dependent threading of an Aharonov-Bohm flux, and the build-up of magnetization at the edges. The magnetization at the inner edge should be measured using an accurate nanoscale scanning magnetometer \cite{Martin1987,Rugar1993,Yacoby2013,Kirtley1995}, depicted by the tip (brown) above the sample. ({\bf b-d}) Evolution of the system during the weak measurement protocol: The system is prepared in its ground state ({\bf b}) and a very weak applied magnetic field induces a small gap, $\Delta_h$. Once the flux is threaded and the levels start climbing/descending according to their spin ({\bf c}). Next a detector which is weakly coupled to the magnetization on the edge is measured ($A$).  For $\Delta_h \rightarrow 0$, the flux threading would typically induce a diabatic Zener-transition ({\bf c}, solid lines) and occasionally no transition ({\bf c}, dashed lines). If the former occurs, the measurement, $A$, is discarded, and if the latter occurs ({\bf d}) it is registered. For a topological insulator the resulting conditional average diverges according to Eq. (\ref{eq-Ags}) \label{fig1}}
\end{figure}

Let us begin with some background on the aforementioned Zener transition. For many purposes, a TI can be thought of as a double layer system wherein one layer consists of only spin-up ($s=1$) electrons and is in an integer quantum Hall effect with a Hall conductance ($\sigma_{xy}$) of $e^2/h$ and the other layer consists of only spin-down ($s=-1$) electrons and is in an integer quantum Hall effect with an opposite Hall conductance \cite{RMP_TI2}. We focus on the inner edge of one layer in a Corbino-disk geometry, initially with no interactions and no disorder, such that the edge conserves the momentum parallel to it ($k_{||}$). One then finds a branch of chiral modes confined to the boundary ($E^{\uparrow}_{k_{||}}$) \cite{Halperin1982}. The sign of the slope of these chiral modes ($\operatorname{sign}[\partial_{k_{||}} E^{\uparrow}_{k_{||}}]$), is determined by the sign of $\sigma_{xy}$, or equivalently in our setup, by $s$. For a finite boundary of length $L$, the allowed momenta along this chiral branch are quantized to $k_{||}(n) = \frac{2\pi (n+\phi/\phi_0) }{L}$ \cite{Half} with $n$ being an integer and $\phi$ is the AB flux through the Corbino disk (see Fig.~\ref{fig1}b, red branch, and imagine that the crossing there is un-avoided). In the many-body ground state, all states with momenta $k_{||}(n)$ such that $E^{\uparrow}_{k_{||}(n)} < \mu$, where $\mu$ is the chemical potential, are occupied (full red circles). 

Using the adiabatic approximation one finds that threading a single flux quantum ($\phi \rightarrow \phi + \phi_0$) changes the many-body ground state to a state in which all $k_{||}(n)$ such that $E^{\uparrow}_{k_{||}(n-1)} < \mu$, are occupied. The latter is an excited state with one additional spin-up electron. Considering the other, spin-down layer, the opposite effect would occur and one spin-down electron would be depleted from the boundary. In total, the number of electron has not changed but a spin-flip excitation was created. Notably the outer edge would exhibits a similar yet opposite effect. For a large system and at low energies compared to the bulk gap, the edges are effectively two decoupled systems \cite{RMP_TI2}. Correspondingly one may ignore the outer edge in all of the following. 

A threading of an AB flux is a periodic cycle in parameter space (up to an insignificant gauge transformation) and the fact that the system does not come back to its ground state after such a cycle necessitates a level crossing. This level crossing occurs at $\phi = \phi_0/2$, when an empty level of the spin-down branch, which is being pushed down in energy by the flux, becomes degenerate with a full spin-up level, being pushed up by the flux. Consequently the excitation, consisting of shifting the electron between these two levels, would cost zero energy. Provided TRS is conserved, no TRS respecting operator can open a gap at this crossing point. This includes operators related to quench disorder and electron-electron interactions. Furthermore, one may naturally extend the definition of TRS to a time dependent Hamiltonians (${\rm H}[t]$) via \cite{Fu2006,Ringel2013} $H[t] = {\rm \Theta} H[-t] {\rm \Theta}^{-1}$, where ${\rm \Theta}= K is_y$, where $K$ denotes complex conjugation and $s_y$ is $y-$Pauli matrix acting on the spin degrees of freedom. This slightly stronger requirement ensures that an orthogonal state is obtained independently of the flux threading rate \cite{Ringel2013,Ringel2014,speed}. 


We begin by describing the measurement setup which consists of a TI and a detector coupled to each other in the presence of a small ambient magnetic field. The respective Hamiltonians are
\begin{align}
{\rm H}[t] ={\rm H}_{TI}[t]+{\rm H}_{pert} + {\rm H}_{detector}\,.
\end{align}
We have in mind a 2D TI in a Corbino disk geometry (see Fig.~\ref{fig1}a). Its Hamiltonian (${\rm H}_{TI}[t]$) depends on the AB-flux, $\phi(t) = \phi_0 \frac{t}{t_f}$, that is generated by a solenoid situated within the hole (see Fig.~\ref{fig1}a). As is, each threading of a $\phi_0$, would induces the aforementioned magnetic excitations on both the inner and outer edges. Applying a small magnetic field $B_i$ at the inner edge introduces the perturbation, $H_{\rm pert}=B_i M_i$, where $M_i$ is the total magnetization on the inner edge in the $\hat{i}$-direction. Generically, this would cause a small gap, $\Delta_h$, between the counter-propagating spin edges (see Fig.~\ref{fig1}b). Below we will always consider the limit $\Delta^{-1} \ll t_f/\hbar \ll \Delta^{-1}_h$ where $\Delta$ is the bulk gap of the TI. 
 
We propose to monitor the system evolution by coupling a weak detector to the magnetization at the inner edge. The detector is modeled as a Harmonic oscillator with a low frequency, $\omega$, whose momentum, $P$, is weakly coupled to the system, namely 
\begin{align}
\label{Eq:HDetector}
{\rm H}_{detector} &= \frac{M \omega^2 X^2}{2} + \frac{P^2}{2 M} +  \lambda P M_i.
\end{align}
Notice however that in the setup suggested in Fig.~\ref{fig1}a $P$ is actually the vertical position operator of the cantilever. Later we comment on how to choose $\lambda$ and calibrate our measurement of $X$.


The measurement protocol begins with both system and detector in their respective ground states, i.e.~the initial state is $|i\rangle = | i_S \rangle | i_D \rangle$. Pictorially, the system's ground state ($|i_S\rangle$) corresponds to Fig.~\ref{fig1}b. Next $\phi(t)$ is scanned, at constant rate, from $0$ to $\phi_0$. Considering the limit of small $\lambda$, one can use first order time dependent perturbation and express the final state ($|f\rangle$) as 
\begin{align}
|f\rangle &= \left(U_0^{t_f} + i \lambda \int^{t_f}_0 \frac{dt}{\hbar} U^{t_f}_t M_z P  U_0^t \right)|i\rangle\,, \\  
U^{t_1}_{t_0} &= \mathcal{T}  e^{i\int^{t_1}_{t_0}\frac{dt}{\hbar} {\rm H}_{TI}[t]+{\rm H_{pert}}}\,, \nonumber
\end{align}
where, for simplicity, we have assumed the time scales of the detector to be much longer than $t_f$, allowing us to ignore the free evolution of the detector. The final state can be readily evaluated on the product basis consisting of detector position basis $\left\{| x \rangle\right\}$ and many-body edge excitation spectrum $\left\{|m\rangle\right\}$, 
\begin{align}
\langle x &|\langle m| | f\rangle &= \langle m |U_0^{t_f}| i_S \rangle \langle x | e^{\lambda \hbar^{-1} A_m P} | i_D \rangle\ + O(\lambda^2),  
\label{Eq:Finalstate}
\end{align}
where
\begin{align}
\label{Eq:Am}
A_m &= i \frac{\langle m | \int dt U^{t_f}_t M_i U_0^t |i_S \rangle}{\langle m | U_0^{t_f} | i_S \rangle} = \hbar\partial_{B_i} \log \left[\langle m | U_0^{t_f} | i_S \rangle \right] \,,
\end{align}
Notably in the last equality we exploited the fact that $B_i$ couples to the same operator as $P$ does. 
The $A_m$'s are known as weak-values (WVs), and for the re-exponentiation ($e^{\lambda \hbar^{-1} A_m P} \approx 1+ \lambda \hbar^{-1} A_m P$) we have assumed that higher-order WVs are negligible \cite{Aharonov1988,Oded2011,dressel2014colloquium}. Since the characteristic scale of $P$ is $\sqrt{\hbar M \omega}$, the condition for a weak measurement is 
\begin{align}
\label{Eq:Weak}
\lambda A_m \ll \frac{1}{\sqrt{ \hbar M \omega} }.
\end{align}


Following the weak measurement protocol \cite{Aharonov1988} one now applies a second strong measurement that determines the final state of the system (postselection) in order to measure a specific WV.  In our case, we consider a postselection on the ground state of the system at time $t_f$ with $\phi =\phi_0$, i.e.~, $|m\rangle \equiv G_{\phi_0} |i_S\rangle$, where $G_{\phi_0}$ is a gauge transformation which inserts a flux quantum through the disk ($G_{\phi_0}=e^{i \theta}$, where $\theta$ is the angle along the disk). The weak measurement outcomes are collected conditional on the postselection outcome, i.e.~, if the system is not found to be in its ground state, the experiment outcome is ignored. Since both the initial and final states are ground states, we denote this weak value simply as $A_{gs}$ from now on.

According to Eq. (\ref{Eq:Finalstate}), the collapse of the system's state on $G_{\phi_0} |i_S\rangle$ leaves the detector in a pure state given by $e^{\lambda \hbar^{-1} A_{gs} P} | i_D \rangle + O(\lambda^2)$. Assuming for the moment that $A_{gs}$ is purely imaginary, the ground state of the detector is simply shifted by $\lambda A_{gs}$ and consequently a standard measurement of $X$ would give $\langle X \rangle_{weak} =  \lambda A_{gs}$. 

Let us turn to evaluate $A_{gs}$. Following Eq. (\ref{Eq:Am}), this amounts to evaluating $\langle i_S| G_{\phi_0}^{\dagger} U_0^{t_f}|i_S\rangle$. Taking the simplest level crossing model and assuming $\Delta_h t_f/\hbar \ll 1$ one may use the well known result \cite{Zener1932,Stueckelberg1932}
\begin{align}
|\langle i_S | G_{\phi_0}^{\dagger}  U_0^{t_f} | i_S \rangle|^2 = c_0 B_i^2 + O(B_i^4),
\end{align}
where $c_0$ is some non-universal constant which depends, in particular, on the rate of flux threading, and the direction of the magnetic field ($\hat{i}$). As previously mentioned, taking a more realistic descriptions of this transition, allowing for example several nearby energy levels, would not alter this result \cite{Ringel2013,Ringel2014}. Comparatively, for a trivial band insulator, with no edge modes, the only energy scale is $\Delta$ with respect to which the experiment is adiabatic and consequently, the above probability changes to $1 - O(B_i^2)$. Plugging these expressions into Eq. (\ref{Eq:Am}), the non-universal contributions decouple and one obtains
\begin{align}
\label{eq-Ags}
A_{gs} &= \frac{2 \hbar \nu_2}{B_i} + O(1) 
\end{align}
where $\nu_2 = 1 (0)$ for a TI (band insulator). Roughly speaking, this follows from viewing the effective magnetic perturbation as $(B_i + \lambda P)$. Postselecting for an avoided Zener transition then means that as $B_i$ decreases, a strong fluctuation in $\lambda P$ is required to assist an avoided transition. 

Notably however, $A_{gs}$ turned out real. Consequently, one extra procedure is needed to witness its effect on the detector position. Re-expressing the detector state following the postselection as 
\begin{align}
e^{\lambda \hbar^{-1} A_{gs} P} &| i_D \rangle = \left(\frac{\hbar}{2\pi M \omega}\right)^{1/2} \int dk  e^{- \frac{\hbar k^2}{2M \omega} + \lambda A_{gs} k} | k \rangle \\ \ \nonumber 
&= \left(\frac{\hbar}{2\pi M \omega}\right)^{1/2} \int dk  e^{- \frac{\hbar \left[k - \lambda \hbar^{-1} A_{gs} M \omega\right] ^2 + O\left(\lambda^2 \right)}{2M \omega}} | k \rangle,
\end{align}
one finds that up to corrections in $\lambda^2$, which we consistently neglect, the weak value simply shifts the momentum ($P$) by $\omega M \lambda \hbar^{-1}A_{gs} $. Waiting for the resulting coherent state of the detector to evolve for a quarter period ($\pi/2$) and then measuring $X$, yields 
\begin{align}
B_i \frac{\langle X \rangle_{weak,\pi/2}}{\lambda} &= 2\hbar \nu_2. 
\end{align}
Equation (\ref{Eq:Weak}) ensures that $\langle X \rangle_{weak,\pi/2}$ is much smaller than the standard deviation of $X$. Notably all the quantities on the l.h.s. are system independent quantities associated with the detector and external perturbation, while the r.h.s is quantized in units of $\hbar$. This is the key result of this work. 

A few comments are in order regarding the observability of the above result. First, obtaining a divergent weak value does not imply that the detector signal actually diverges, as that would mean that the measurement ceases to be weak (see Eq. (\ref{Eq:Weak})). Instead as $A_{gs}$ diverges, $\lambda$ must be reduced and the detector read-out must be recalibrated using an independent classical source of magnetization. Alternatively stated, we treat $\frac{\langle X \rangle_{weak,\pi/2}}{\lambda}$ as the calibrated detector read-out and only in terms of this value would the signal appear divergent. 

Second, the quantization depends on an accurate detection of the final state. Final state detection errors would cut-off the divergent nature of a weak value \cite{Romito2012}. Since we require single spin levels of detection, such errors are unavoidable with current technology although the field is progressing rapidly \cite{Yacoby2013}. In this aspect, it would be beneficial to choose the axis along which the final state magnetization is measured parallel to the anticipated direction of magnetization. 


Third, an error in the quantization of the pole's residue would be induced by tilting the direction of perturbing field ($B_{i}$) with respect to that of the magnetization being measured ($M_i$). Less restrictively, all is required is that the TRS breaking perturbation couples to same TRS breaking operator which is being weakly measured. This operator can be any TRS breaking operator and, in particular, may vary in space. To achieve such coupling, one may use an invasive magnetometer, such as a magnetized AFM tip, to both generate the perturbation and measure it. 

Fourth, since in practice one cannot take the limit of infinitely weak $\Delta_h$ (and therefore infinitely weak $\lambda$), almost unavoided Zener transitions may also contribute to the pole's residue in a non-generic way. However away from fine-tuned points and for small hole circumference, we do not expect such near degeneracies neither for TIs nor for trivial insulators. 


Lastly, if $\langle X \rangle_{weak}$ is to contain any information about the final magnetization measurement the operators $M_i$ carried by $U_t^{t_f}$ up to time $t_f$, should not commute those measuring the final state \cite{Romito2008}. For a TI in which charge is the only conserved quantity, this is generically the case. However given extra symmetries, for example $s_z$ conservation, the axis of the final magnetization measurement and $M_i$ must not be both aligned along $\hat{z}$. These requirements were implicit in our treatment via the assumptions that the perturbation induces a finite $\Delta_h$ gap and that the final measurement distinguishes the ground state from the excited state. 

The quantized residue obtained for the TI, relied mainly on having an unavoided Zener transition and detector-perturbation alignment. Consequently, it may be observed also in different setups. For instance, one may consider a gate-voltage driven Zener transition between two charge states of a weakly coupled, spin polarized, double quantum dot \cite{Romito2012}. These two states being one electron on the left dot and no electrons on the right dot, and vice versa. Obtaining residue quantization in this setup requires a detector which couples to a tunneling operator (${\rm T}$) that transfers charge between the two dots. Formally this requires ${\rm H}_{detector}$ with $M_i$ replaced by ${\rm T}$. Furthermore, one should control the average value of the detector's momentum ($\langle P \rangle = \lambda^{-1} B_i$) as this effectively generates the analogous term to ${\rm H}_{pert}$. Potentially, such a detector could be realized using an LC circuit whose charging ($P$ in our notations) controls the opening of a quantum point contact (QPC) between the two dots.  Postselection must again be done for the unlikely outcome, being that the electron hopped between the dots.  

Applying the analysis carried earlier in the above mesoscopic setup, the following dependence of the weak value of the flux on the inductor $\langle \phi \rangle_{weak}$ on the average charging bias of the capacitor ($Q_0$) is obtained 
\begin{align}
\langle \phi \rangle_{weak} &= \frac{2 \hbar}{Q_0},
\end{align}
where the weak measurement limit simply requires $Q_0$ to be larger than the zero-point quantum fluctuations of the charging of the capacitor. Of course being related to voltage driven charge transition, rather than an AB flux driven magnetic transition as before, this WV carries no topological meaning. Its also worth noting that a single transition of the latter type cannot be realized in a charge conserving system, although pairs of such transitions may occur \cite{Gefen1987}. This is due to the change time in reversal polarization as a function of the AB flux \cite{Fu2006}. Considering achievable resonator frequencies  \cite{Johansson2006,Hofheinz2008} of $\omega_{LC} \approx 5Ghz $, a quantum coherent weak measurement requires $t_f^{-1} \gg \omega_{LC} \approx 5Ghz \gg k_b T/\hbar$ with $t_f$ limited by the gap to higher energy levels which is controlled by the magnetic field. At one Tesla this implies $t_f^{-1} \ll 100Ghz$. Initial and final state read-out can be achieved using single electron detectors such as an RF-SET \cite{Schoelkopf1998}, or one may weakly couple the dots to leads and deduce their charge from Coulomb blockade peaks in the conductance \cite{livermore1996}. 
 
To conclude, a weak measurement was proposed which identifies Zener transitions through the residue associated with a divergent WV. Interestingly, this residue is quantized in units of $\hbar$. Since a defining property of a TI is the presence of a single flux-driven Zener transition between a Kramer's pair \cite{Fu2006}, this residue yields the $Z_2$ topological index. A different Zener transition, driven by gate voltage, may be realized in mesoscopic setup yielding similar results. This latter setup appears feasible already with current technology.  


{\it Acknowledgments:} Z. R. would like to thank Oded Zilberberg, Peter Leek, and Steven H. Simon for helpful comments and discussions. This work was supported by EPSRC grant no. EP/I032487/1

 \bibliography{WVsTop}

\begin{thebibliography}{36}
\bibitem[{mag()}]{magnetization}
\bibinfo{note}{In the simplistic model described below, this magnetization has
  a magnitude of two spins and points in the $\hat{z}$ direction. Generically
  it may point in any direction, and should typically be of the order of, but
  smaller than, two spins}.
\bibitem[{spe()}]{speed}
\bibinfo{note}{The rate at which flux quanta are threaded must be smaller than
  the bulk gap over $\hbar$. From Ref. ([7]) one obtains a threshold of roughly
  10 Terahertz.}
\bibitem[{man()}]{many}
\bibinfo{note}{At fine tuned points unprotected Zener transition could appear,
  however must do so in pairs. Perturbing the system slightly away from such
  point these excess transitions are removed.}
\bibitem[{Hal()}]{Half}
\bibinfo{note}{More generally one should replace $\phi \rightarrow \phi +
  \phi_{micro}$, where $\phi_{micro}$ is some flux which depends on
  microscopics. Since it bears no consequence on our discussion, it is
  henceforth omitted.}
\end{thebibliography}

\end{document}